# Optimum Transmission Policies for Battery Limited Energy Harvesting Nodes

Kaya Tutuncuoglu      Aylin Yener

Wireless Communications and Networking Laboratory (WCAN)

Electrical Engineering Department

The Pennsylvania State University, University Park, PA 16802

*kaya@psu.edu*      *yener@ee.psu.edu*

## Abstract

Wireless networks with energy harvesting battery equipped nodes are quickly emerging as a viable option for future wireless networks with extended lifetime. Equally important to their counterpart in the design of energy harvesting radios are the design principles that this new networking paradigm calls for. In particular, unlike wireless networks considered to date, the energy replenishment process and the storage constraints of the rechargeable batteries need to be taken into account in designing efficient transmission strategies. In this work, such transmission policies for rechargeable nodes are considered, and optimum solutions for two related problems are identified. Specifically, the transmission policy that maximizes the short term throughput, i.e., the amount of data transmitted in a finite time horizon is found. In addition, the relation of this optimization problem to another, namely, the minimization of the transmission completion time for a given amount of data is demonstrated, which leads to the solution of the latter as well. The optimum transmission policies are identified under the constraints on energy causality, i.e., energy replenishment process, as well as the energy storage, i.e., battery capacity. For battery replenishment, a model with discrete packets of energy arrivals is considered. The necessary conditions that the throughput-optimal allocation satisfies are derived, and then the algorithm that finds the optimal transmission policy with respect to the short-term throughput and the minimum transmission completion time is given. Numerical results are presented to confirm the analytical findings.

This work was supported by NSF Grant CNS 0964364. The conference paper [1], scheduled to appear at ICC 2011, contains technical material in part from the short term throughput optimization problem considered in this submission.

# I. INTRODUCTION

With the wide deployment of battery powered wireless devices, prolonging the lifetime of wireless networks is becoming ever more critical [2]. Systems powered with batteries suffer from a limited lifetime, whereas networks consisting of energy harvesting or *rechargeable* nodes can survive perpetually [3]. The performance that such systems can deliver is tied closely to efficient utilization of energy that is currently available, as well as that is to be harvested. Towards understanding fundamental performance limits with energy harvesting, in this work, we identify optimal transmission policies for maximizing throughput and minimizing the transmission completion time of a wireless node with a replenishing energy source and a finite battery.

There has been a significant amount of previous work in energy efficient communications and networking of nodes with non-rechargeable batteries, from the perspective of various protocol layers, or cross-layer approaches, e.g. [4]–[12]. References relevant to this paper include [4], [5]. Energy efficient deadline-constrained communications is considered in [4] resulting in a packet scheduling algorithm that achieves minimum energy. In [5], an insightful approach to energy-efficient rate control is developed and the problem with individual deadline or buffer constrained systems with data arrivals is solved.

There has also been some recent interest in wireless networks with energy harvesting nodes. A queue stabilizing transmission policy is developed in [13] for a recharging battery powered transmitter. This is a modified adaptive backpressure policy that is shown to be asymptotically optimal for sufficiently large battery capacity. Reference [14] calculates energy management policies that are throughput optimal or mean delay optimal, and shows that a greedy policy is optimal for both problems in low SNR regime. A discrete-time battery model is introduced in [15] for the problem of energy aware routing in wireless networks powered with renewable energy. Reference [16] assumes a Markov model for battery state, and proposes a threshold algorithm to best utilize available energy to packets with different rewards. In [17], policies based on the energy-error probability tradeoff are developed to maximize successful transmission probability while minimizing probability of running out of energy in an energy harvesting body sensor network. For a solar powered wireless network, [18] lays out sleep/wake-up strategies for various factors and determines optimal parameters of the solar energy harvest based strategy using a bargaining game model. The most relevant reference work to the present paper is [19], [20], where the authors have considered an energy harvesting node, and found transmission policies that minimize the transmission completion time of a given amount of data. This work assumes an infinite capacity battery for the energy harvesting transmitter.

In this paper, we consider the problem of maximizing *the short-term throughput* of an energy harvesting node, in other words, maximizing the data transferred under a deadline constraint. Unlike previous work [19], we employ the realistic constraint that an energy harvesting battery must have finite energy storage capacity. We seek to find the optimum transmission policy under this energy storage constraint as well as the energy causality constraint. That is to say that, due to the finite battery at the transmitter, any received energy that overflows its capacity is lost, and, that energy can not be expended prior to being harvested. Like previous work [19], we consider known energy arrivals, in order to find a bench mark solution for any transmission policy that considers communication under a deadline. We also show that the problem solved in this paper is closely related to the problem of transmission completion time minimization considered in [19]. Specifically, we show that the solution to the former is identical to that of the latter for the same parameters, providing a solution to the latter under battery storage constraints as well. The algorithm that finds the optimum power allocation to solve the former is developed. Then, using the relation between the two problems, a similar algorithm that finds the optimum power allocation to minimize the transmission completion time is presented.

The remainder of the paper is organized as follows. The system model and the problem definition is presented in Section II. Section III solves the short-term throughput maximization problem and presents the optimal policy. Section IV addresses the completion time optimization problem. Numerical results are presented in Section V followed by concluding remarks in Section VI.

## II. System Model and Problem Definition

We assume a single link continuous time system where a node transmits continuously and its rate can be varied at will via power control. Specifically, the transmitter can choose to transmit with power $P(t)$ at any instant $t$, achieving a corresponding rate $r(p(t))$ where $r(.)$ is a non-negative, increasing, strictly concave function that we will refer to as the *power-rate function*. Using a power-rate function of this form is fairly common [5], [20], and is clearly valid for the additive white Gaussian noise (AWGN) channel. Factors such as processing power, battery leakage or base energy consumption are not explicitly taken into account in this model, however, can be integrated in the power-rate function without violating its above stated properties. For example, a base energy consumption of power $P_b$ could be modeled as a shift in the power rate function horizontally by $P_b$, or a processing power linear in transmission rate could be added to $r(.)$ to yield a new power-rate function satisfying the same properties.

We consider an energy harvesting system with a finite-capacity battery. That is to say that the battery can store up to $E_{max}$. This storage capacity is considered to be constant throughout the transmission horizon, i.e., battery wear or fatigue is assumed to be much slower than the time scale of the problem. The property of keeping the battery energy below its capacity and non-negative, i.e. within $[0, E_{max}]$, will be referred to as *energy-feasibility* in the sequel.

We will consider the offline optimization problem. Hence the energy replenishment process is modeled as a discrete process with energy harvests of size $E_n$ arriving at time instances $s_n$, as shown in Figure 1, where $E_n$ and $s_n$ take positive real values and are available non-causally to the transmitter at the beginning of transmission. The first energy arrival $E_0$ is conventionally at time $s_0 = 0$, representing the initial energy in the battery. Due to the limited battery, if the harvested energy $E_n$ is larger than the available space in the battery at time $s_n$, the battery is charged to maximum capacity and the remainder of the energy packet is discarded. Note that an instantaneous energy consumption requires infinite instantaneous power, which is not allowed by system definition. Thus, all harvested energy must first be stored before consumption. Based on this observation, it is safe to assume that $E_n \leq E_{max}$, as no more than an $E_{max}$ amount of a harvest can be utilized by the transmitter. Therefore, a truncation at $E_{max}$ for $E_n$ values will be assumed in the sequel.

An example to justify this model could be a solar powered surveillance network. A continuous operation would be necessary for surveillance purposes, and the nodes would realistically be operating on finite batteries. Once the nodes are deployed, the solar energy harvesting process would be predictable enough to have good estimates of $E_n$ and $s_n$, yet significantly varying throughout the day.

In this model, the amount of energy available at any time instant is constrained, either because a sufficient amount may not yet be harvested or cannot be stored in the battery. As such, we have an energy-feasibility constraint on the transmission policy. Specifically, a *energy-feasible* power allocation $p(t)$ is a bounded non-negative function for the transmission power that ensures the battery state stays within $[0, E_{max}]$. We express the set of energy-feasible power allocation functions as

$$\mathfrak{P} = \left\{ \, p(t) \mid 0 \leq \sum_{k=0}^{n-1} E_k - \int_0^{t'} p(t)\mathrm{dt} \leq E_{max}, \ s_{n-1} \leq t' \leq s_n \, \right\} \tag{1}$$

which is a convex set of functions, i.e., any convex combination of two energy feasible allocations is also energy feasible. Note that allocation functions which lead to a battery overflow are physically possible, but are discarded from the energy-feasible set due to being strictly suboptimal as shown in subsequent sections of this paper.

Our goal is to find the optimum power allocation for maximizing the total data transferred from an energy harvesting node under a deadline constraint. The only constraint induced by the model on the power allocation is the feasibility set $\mathfrak{P}$. The objective is to maximize the total number of bits departed in the time interval $[0, T]$ over $p(t)$. Under a rate function $r(p(t))$ and a time deadline $T$ the problem can be expressed as:

$$\mathbf{P1}: \quad \max_{p(t)} \ \int_0^T r(p(t))\mathrm{dt}, \qquad \text{s.t.} \quad p(t) \in \mathfrak{P} \tag{2}$$

Next, we shall solve this problem.

## III. SHORT-TERM THROUGHPUT MAXIMIZATION

This section considers the throughput maximization problem for an energy harvesting node with infinite backlog and finite time. First, the necessary properties of the optimal policy are established. Then, an algorithm is presented to generate the policy that satisfies all the necessary conditions and the optimality of this policy is shown.

### A. Optimality conditions

The following lemmas provide the necessary conditions for a power allocation policy to be optimal. These conditions provide valuable insight into the development of the algorithm proposed in Section III-B and the proof of its optimality.

*Lemma 1:* Given the total amount of energy consumed in any time interval $[t_1, t_2]$, a constant-power transmission is throughput optimal, i.e., among all transmission policies consuming energy $E$, the constant-power transmission $p(t) = \frac{E}{t_2 - t_1}$ departs the largest number of bits.

*Proof:* The proof is by contradiction. Assume that $p(t)$ is non-uniform in $(t_1, t_2)$ , i.e., for $\tau_1, \tau_2 \in (t_1, t_2)$, $\tau_1 \neq \tau_2$, we have $p(\tau_1) \neq p(\tau_2)$. Without loss of generality, assume that $p(\tau_1) < p(\tau_2)$. Keeping $p(t)$ unchanged at every other time instance, we define an alternative policy $p'(t)$ as

$$p'(t) = \begin{cases} p(t) + \delta & t \in [\tau_1, \tau_1 + \epsilon] \\ p(t) - \delta & t \in [\tau_2, \tau_2 + \epsilon] \\ p(t) & otherwise. \end{cases} \quad (3)$$

where $\delta > 0$ is an infinitesimal power displacement among two intervals of arbitrarily small duration $\epsilon > 0$. Clearly, the energy consumed by $p(t)$ and $p'(t)$ are equal. However calculating the information sent in $[t_1, t_2]$ yields

$$B = \int_{t_1}^{t_2} r(p(t)) \; dt$$
$$= \int_{t_1}^{\tau_1} r(p(t)) \; dt + \int_{\tau_1 + \epsilon}^{\tau_2} r(p(t)) \; dt + \int_{\tau_2 + \epsilon}^{t_2} r(p(t)) \; dt + [r(p(\tau_1)) + r(p(\tau_2))].\epsilon$$

$$< \int_{t_1}^{\tau_1} r(p(t)) \ dt + \int_{\tau_1+\epsilon}^{\tau_2} r(p(t)) \ dt + \int_{\tau_2+\epsilon}^{t_2} r(p(t)) \ dt + (r(p(\tau_1)+\delta) + r(p(\tau_2)-\delta)).\epsilon$$

$$\tag{4}$$

$$= \int_{t_1}^{t_2} r(p'(t)) \ dt = B'$$

where the inequality in (4) follows from the definition of strict concavity for $r(p)$. This shows that we can strictly improve the throughput by equalizing the two power values, and by extension uniformizing the power allocation. ∎

*Corollary 1:* The optimal power allocation policy dictates the transmission power, and thus the rate, remains constant between energy arrivals. Consequently the power level can change *only* when a new energy packet arrives. This follows from Lemma 1, by considering the interval $[s_n, s_{n+1}]$ and introducing the feasibility definition in (1) to the optimization. However, given a total consumed energy in $[s_n, s_{n+1}]$, feasibility in the sense defined in (1) does not depend on the structure of $p(t)$ in this interval. Thus the constant power policy is feasible, and therefore optimal for any given energy consumed.

*Lemma 2:* Any power allocation policy yielding a battery overflow is strictly suboptimal.

*Proof:* It is clear that the battery cannot possibly overflow without an energy arrival. Therefore, we start by assuming that the battery overflows at the energy arrival instant $s_i$. Let the power allocation allowing this overflow be $p(t)$. $E_i$ is less than or equal to $E_{max}$ by system model and causes an overflow, thus battery state at $s_i^-$ is strictly positive. This implies that $p(t)$ can be increased by an infinitesimal amount $\delta$ in $(s_i - \epsilon, s_i)$ without violating energy-feasibility, which strictly increases the throughput. Since the excess energy required, $\delta.\epsilon$, is recovered by the overflow at $s_i$, the remaining transmission schedule remains unchanged. Therefore, a power allocation that yields a battery overflow can not be optimal. ∎

*Lemma 3:* In the optimal power allocation, the transmission power does not change unless the battery is either full or completely depleted.

*Proof:* Assume that at arbitrary time $t$ transmission power changes, so that $p(t^-) \neq p(t^+)$. Consider the interval $[t - \tau, t + \tau]$, where the policy depletes a total energy of $\tau.(p(t^-) + p(t^+))$. Unless the battery is full or depleted at $t$, the energy feasibility constraints will be inactive in this interval. Let $p^*(t) = \frac{p(t^-) + p(t^+)}{2}$ be the constant power transmission policy in $[t - \tau, t + \tau]$ that expends the same total energy, which is also feasible for $\tau$ sufficiently small, due to energy feasibility constraints being locally inactive around $t$. Then, $p(t)$ can be replaced with $p^*(t)$ without altering the rest of the problem, and due to Lemma 1, departs strictly more bits. Thus, $p(t)$ must be stay constant unless an energy feasibility constraint is active, i.e. the battery is either full or depleted. ∎

Note that in the proof above, $p^*(t)$ might still be feasible when battery is full or depleted, specifically if it shifts the policy away from the active constraint. This observation forms the grounds of the next lemma:

*Lemma 4:* For optimal power allocation, the change in power level $p_n$ at an energy arrival instant $s_n$ has to be nonnegative (nonpositive) if the battery is depleted (full) at that time instant.

*Proof:* The proof is by contradiction. Consider the notation in the proof of Lemma 3. Battery being depleted at time $t$ implies that the first inequality in (1) is active. However, if $p(t^-) > p(t^+)$ holds, i.e. if transmission power is decreasing, then $p^*(t) < p(t)$. Hence replacing $p(t)$ with $p^*(t)$ on $[t - \tau, t + \tau]$ increases the RHS of the active inequality, which is feasible. This shows that if the battery is depleted at $t$, a decreasing power policy at $t$ can not be optimal. Similarly, battery being full at time $t$ implies that the second inequality in (1) is active. This time, an increasing power policy, $p(t^-) < p(t^+)$, implies $p^*(t) > p(t)$; which renders replacing $p(t)$ with $p^*(t)$ on $[t - \tau, t + \tau]$ feasible. This proves the second statement of the lemma. ∎

*Corollary 2:* Lemmas 3 and 4 together imply that for optimal power allocation, transmission power decreases only at energy arrival instants when the battery is full and increases only at energy arrival instants when the battery is depleted.

*Lemma 5:* The optimal power allocation expends all harvested energy by the end of the transmission.

*Proof:* Assume that in an energy-feasible scenario, there is energy left after the transmission concludes. Without violating feasibility, this energy can be used in a nonzero time interval right before the end of transmission. This, due to power-rate function being increasing, strictly improves the throughput. Therefore the former policy can not be optimal. ∎

### B. Throughput Maximizing Policy

First, we present the definitions of the variables used in the policy. From Lemmas 1-4 in section III-A, we know that the power allocation policy will consist of intervals of constant power, changing at some energy harvest instances $s_n$ only. We denote the subsequence of $s_n$ at which the transmission power changes as $i_n$. Let $s_{n_{max}} = T$ be a dummy harvest indicating the end of transmission. The energy of this dummy harvest is arbitrary as it cannot be used by the transmitter, and is set to $E_{n_{max}} = E_{max}$ for convenience. The power allocation then has to be of the form

$$p(t) = \begin{cases} p_n & i_{n-1} < t < i_n \\ 0 & t > s_{n_{max}} = T \end{cases} \tag{5}$$

We remark that once the specifics of the interval $t = [0, i_1)$ is determined, the remainder of the problem can be considered as a separate throughput maximization problem. That is, given the duration of this interval $i_1$, and the amount of information sent in this epoch, $i_1.r(p_1)$, it remains to solve for the optimal power allocation for a modified problem with energy arrival times shifted by $i_1$, a new initial battery state $\sum_{k:s_k < i_1} E_k - i_1.p_1$, and a new deadline constraint of $T - i_1$. This means that once the first time slot of the optimal allocation is identified, the remaining power levels can be found recursively with the same algorithm, using updated parameters and shifted arrival times. Thus, we shall focus on determining the optimal power level in the initial epoch. The modified optimization problem described above will be referred to as the *shifted* optimization problem.

We define two sets of powers $\{p_0[1], p_0[2], ...\}$ and $\{p_{max}[1], p_{max}[2], ...\}$, where $p_0[n]$ and $p_{max}[n]$ are the constant power levels that would result in an empty battery at $s_n^-$ or a full battery at $s_n^+$ respectively if employed in $[0, s_n]$. Note that these levels need not be feasible, or even positive, but only serve comparison purposes. We then define the set $\mathbf{P} = \{\mathbf{P}[1], \mathbf{P}[2], ...\}$ with elements as the closed intervals $\mathbf{P}[n] = [p_{max}[n], p_0[n]]$ between corresponding elements of the two sets $p_0$ and $p_{max}$. This translates to a range of constant power levels for the $n^{th}$ arrival that would be energy-feasible at $s_n$ when the feasibility at previous arrivals are disregarded. Thus, we have:

$$p_0[n] = \frac{\sum_{k=0}^{n-1} E_k}{s_n}, \quad p_{max}[n] = \frac{\sum_{k=0}^{n} E_k - E_{max}}{s_n} \qquad\qquad s_n < T \qquad (6)$$

$$\mathbf{P}[n] = [p_{max}[n], p_0[n]] = \{p \mid p_{max}[n] \leq p \leq p_0[n]\} \qquad\qquad s_n < T \qquad (7)$$

$$\mathbf{P}[n_{max}] = \{p_0[n_{max}]\} \qquad\qquad s_{n_{max}} = T \qquad (8)$$

Here, (8) follows due to $E_{n_{max}} = E_{max}$ yielding $p_0[n_{max}] = p_{max}[n_{max}]$. Since Lemma 5 readily suggests that the energy should be depleted at $s_{n_{max}}$, the convenient choice of $E_{n_{max}}$ is justified. Based on this definition of the feasible power range, it can be deduced that for a constant power transmission starting from $t = 0$ to extend to the $n^{th}$ energy harvest without violating energy-feasibility, its power level should be contained in the range $\mathbf{P}[k]$ for $k = 1, ..., n$, i.e., the step should be feasible through all harvests it extends over. This yields an upper bound $n_{ub}$ on the length of the first constant power transmission, that can be calculated as

$$n_{ub} = \max\{n \mid \bigcap_{k=1}^{n} \mathbf{P}[k] \neq \varnothing, n = 1, 2, .., n_{max}\} \qquad (9)$$

as for the later harvests, a feasible constant power level contained in all previous feasible power ranges $\mathbf{P}[n]$ do not exist. This gives us a range for feasible constant power transmission levels and its maximum duration.

*Remark 1:* The following observation explains the intuition behind the optimal algorithm that will be proposed. Given an energy harvesting scenario and the corresponding $\{p_0[n]\}$ and $\{p_{max}[n]\}$ sets, assume that a constant transmission of power $p_1$ and duration $s_{n_1}$ is feasible. This transmission then satisfies $p_1 \in \bigcap_{k=1}^{n_1} \mathbf{P}[k]$ and cannot extend beyond $s_{n_{ub}}$, as it is rendered infeasible at $s_{n_{ub}+1}$ by one of the constraints. However, which constraint causes this is an important indication of how the policy tends to change after $s_{n_1}$.

A constant transmission with power $p_1$ either over-depletes or overflows the battery at $s_{n_{ub}+1}$. The former case implies that the power level after $s_{n_1}$ needs to decrease, and the latter implies that the power level needs to increase in order to conform to feasibility conditions at $s_{n_{ub}}$. This can be verified by calculating updated values of $p_0[n]$ and $p_{max}[n]$ for a shifted problem after the first step of the policy is determined. Figure 2 shows how the boundaries of the interval $\mathbf{P}[n]$ for the shifted problem, shown in red, move away from the chosen transmission power $p_1$ when re-calculated starting from time $s_{n_1}$. This indicates that if an interval $\mathbf{P}[n]$ falls below or above a chosen transmission power, than the corresponding intervals of the shifted problem will again fall above and below respectively, forcing the power of the next step to increase and decrease respectively. By Corollary 2, an increase or decrease in power can occur only at a harvest point and with the battery depleted or full. Hence, the choice of $p_1$ in the optimal policy is restricted to $p_0[n_1]$ and $p_{max}[n_1]$ respectively for the two cases in consideration. Observing the tendency of the power level to increase or decrease after $s_{n_1}$ therefore determines the power level $p_1$.

The analysis and intuition presented so far aims at insights on the optimal power allocation. Based on these, we prove that the following algorithm determines the throughput maximizing power allocation for a transmitter node with initial energy $E_0$, energy arrivals $E_n$ at times $s_n$, battery capacity $E_{max}$ and a deadline constraint $T$:

### *Throughput Maximizing Algorithm, A1*

1) Find the upper bound on the length of the first constant transmission using 9. If $n_{ub} = n_{max}$, transmit with constant power $(\sum_{k=0}^{n_{max}} E_k)/T$ until the end of transmission.

2) Determine whether the next power interval $\mathbf{P}[n_{ub} + 1]$ falls below or above $\bigcap_{k=0}^{n_{ub}} \mathbf{P}[k]$.

3) If $\mathbf{P}[n_{ub} + 1] > \bigcap_{k=0}^{n_{ub}} \mathbf{P}[k]$, transmit with

$$i_1 = s_{n_1}, \quad p_1 = p_0[n_1], \quad \text{where } n_1 = \max\{n \mid p_0[n] \in \bigcap_{k=0}^{n} \mathbf{P}[k]\},$$

If $\mathbf{P}[n_{ub} + 1] < \bigcap_{k=0}^{n_{ub}} \mathbf{P}[k]$, transmit with

$$i_1 = s_{n_1}, \quad p_1 = p_{max}[n_1], \quad \text{where } n_1 = \max\{n \mid p_{max}[n] \in \bigcap_{k=0}^{n} \mathbf{P}[k]\}.$$

4) Repeat algorithm for the shifted problem with modified parameters

$$E'_0 = \sum_{k=0}^{n_1} E_k - i_1.p_1, \quad T' = T - i_1, \quad n'_{max} = n_{max} - n_1,$$

$$s'_n = s_{n+n_1} - i_1, \qquad E'_n = E_{n+n_1}, \qquad n = 1, ..., n'_{max}. \tag{10}$$

A graphical description of the algorithm is provided in Figure 3 which shows the *feasible energy tunnel* of the energy harvesting transmitter. The upper solid boundary represents the cumulative energy harvested and the lower solid boundary is the upper boundary shifted down by an amount of $E_{max}$. The cumulative energy spent by the power allocation algorithm forms a continuous line, and must stay within this tunnel to conform to energy feasibility. A power allocation that goes above the tunnel spends more energy than available, and one that goes below causes a battery overflow. Therefore, the set of energy-feasible power allocations $\mathfrak{P}$ lies within this tunnel. Note that this graphical depiction is similar to that in [5], where a tunnel is formed for the data transferred. In contrast, here, we have an energy tunnel.

In Figure 3, the sets $p_0[n]$ and $p_{max}[n]$ correspond to the slopes of lines from the origin to each of the corner points in the tunnel as shown with dashed lines. The interval $\mathbf{P}[n]$ represents the slope set of lines passing through the $n^{th}$ opening in the tunnel, and is marked with an arc on the figure for $n = 2$. The first step of the algorithm determines the *longest* constant power transmission that stays within this tunnel. The second step determines whether the most distant point on a wall that a line through the origin can reach is an upper bound or a lower bound. This is accomplished by comparing the first unreachable opening with the last reachable one. Finally, the third step selects the longest feasible constant transmission that ends in one of the sets $p_0[n]$ and $p_{max}[n]$, allowing a change in transmission power for the rest of the problem.

*Theorem 1:* Algorithm A1 yields the optimal power allocation policy.

*Proof:* Proof is by contradiction. We begin by assuming that there exists an optimal power allocation step $\{\bar{p_1}\}, \{\bar{i_1}\}$ such that $p_1 \neq \bar{p_1}$ or $i_1 \leq \bar{i_1}$. There are four possible distinct cases over which we will show contradictions on optimality, labeled in Figure 4 as I-IV.

Figure 4(a) shows a case where the algorithm chooses to transmit with power $p_1 = p_{max}[n_1]$

for a duration $i_1 = s_{n_1}$. This selection implies that point A lies on the upper tunnel boundary due to the outcome of the second step of A1. A power allocation policy can differ from this selection by either trespassing to the upper region I or the lower region II. A feasible power allocation forms a continuous line that stays within the tunnel throughout the transmission time. Therefore, a power allocation extending to region II must cross the algorithm's line again before $s_{n_1}$. At this crossing point, the algorithm's transmission departs strictly more bits than the alternative due to Lemma 1, rendering any allocation extending to region II suboptimal. On the other hand, any policy extending to region I must cross the dashed line to be feasible, at which point a constant power transmission up to that point is feasible and performs strictly better. Therefore, any allocation extending to region I cannot be optimal.

A parallel statement holds for the second case shown in Figure 4(b) where the algorithm chooses to transmit with power $p_1 = p_0[n_1]$ instead. Similarly, due to the second step of algorithm, point B is known to lie on a lower boundary. A policy extending to region III has to cross the algorithms line at some point in order to complete a feasible path to the deadline. At this crossing point it proves itself suboptimal as the constant power transmission is strictly better. On the other hand a policy extending to region IV has to cross the dashed line, at which point a constant power transmission up to that point is both feasible and strictly better. Therefore no other policy can perform better than the constant power transmission shown with the solid line.

The termination step follows an analogous logic. Due to Lemma 5, it is known that the policy must terminate with a completely depleted battery at deadline T. The termination step suggests a constant power transmission to this point should be employed whenever feasible. As the constant power transmission is the optimal solution for the unconstrained problem as in Lemma 1, it emerges as the solution to the energy constrained problem as well, provided that it is feasible. Consequently any alternative power allocation departing from this constant power transmission would be suboptimal, as the two policies have to terminate at the same point at which the constant power transmission clearly departs more bits.

In summary, any transmission policy that differs from the one found by the algorithm and still is energy feasible must be suboptimal. Consequently, our proposed algorithm's step has to yield the optimal policy. ■

With the necessary conditions and algorithm A1 derived for the throughput maximization problem, we move on to an alternative problem with similar results.

## IV. TRANSMISSION COMPLETION TIME MINIMIZATION PROBLEM

A previous problem studied with an energy harvesting transmitter is the transmission completion time minimization problem, introduced in [20]. The transmission completion time minimization problem focuses on determining an energy feasible power allocation $p(t) \in \mathfrak{P}$ that, given the total number of bits to send as $B$, finalizes the transmission in the shortest time possible. This problem has total transmission time $T$ as an objective function to be minimized, and a throughput constraint as seen in the mathematical expression of the optimization problem:

$$\mathbf{P2}: \quad \min \quad T, \quad \text{s.t.} \quad B - \int_0^T r(p(t)) \mathrm{d}t \leq 0, \quad p(t) \in \mathfrak{P} \tag{11}$$

In contrast to its throughput maximization counterpart we considered in Section III, the time minimization problem appears to have a more complex form. Reference [20] solved this problem by employing lemmas similar to the lemmas in Section III-A, without the battery constraints. To address the model with finite battery capacity, an extension of the algorithm in [20] is possible. However, we will choose another direction that is simpler.

Specifically, we note that although the throughput maximization problem differs significantly in structure from the transmission completion time minimization problem, their solutions are closely related. Theorem 2 states this relationship.

*Theorem 2:* For a given energy harvesting scenario, the two optimization problems yield identical power allocation policies for matching time and bit constraints. In other words, if the maximum-throughput policy for time interval $[0, T]$ departs a total of $B$ bits, then the minimum-time policy for $B$ bits completes the transmission at time $T$, and vice versa.

*Proof:* First the Lagrangian dual problem of the time minimization problem in (11) is formulated in (12). The variables $T$ and $p(t)$ of the nested minimization problem are by definition independent. Keeping in mind that the solution to the dual problem satisfies $u \geq 0$, the inner problem can be separated as in (13). It can now be readily observed that the optimal power allocation function $p^*(t)$ arises from the solution of the maximization problem marked with $P1$ for the optimal completion time $T^*$. Observe that $P1$ is identical to the throughput maximization problem (2). Therefore the solution of the completion time minimization problem is identical to the solution of the throughput maximization problem where the time constraint is the minimum transmission completion time $T^*$.

$$\max_{u \geq 0} \quad \Big( \min_{p(t) \in \mathfrak{P}, \forall T} \quad \big( T + u(B - \int_0^T r(p(t)\ )\mathrm{d}t)\ \big) \Big) \tag{12}$$

$$\max_{u \geq 0} \quad \Big( \min_{\forall T} \ \big( T + u.B - u. \underbrace{\max_{p(t) \in \mathfrak{P}} \int_0^T r(p(t))\mathrm{d}t\ }_{P1} \big) \Big) \tag{13}$$

∎

*Remark 2:* An alternative proof is by making use of the monotonicity of the two problems in time and bits. The minimum-time problem yields a strictly larger completion time for more bits, while the maximum-throughput problem departs strictly more bits for a later deadline due to the strict concavity of the power-rate function. Assume that the maximum-throughput policy for interval $[0, T]$ departs a total of $B$ bits but the minimum-time policy for $B$ bits at time $0$ completes the transmission at time $T' \neq T$. Consider the two cases $T' > T$ and $T' < T$. The contradiction in the first case $T' > T$ is trivial as the maximum-throughput policy departs the same number of bits in a shorter time; and thus the minimum-time policy cannot be optimal. In the second case, minimum-time policy achieves the same throughput in a shorter time $T' < T$ indicating that there exists a shorter policy with the same throughput. Therefore strictly more than $B$ bits can be sent in the larger time interval $[0, T]$ and thus the suggested maximum-throughput policy cannot be optimal.

Theorem 2 states that the solution to the former and latter problems are in fact identical. Therefore Lemmas 1-5 that characterize the optimal power allocation also apply to the completion time minimization problem. We make use of this relationship to develop a modified algorithm that yields a throughput maximizing power allocation policy while departing exactly the desired number of bits, thus solving the completion time minimization problem.

*Remark 3:* The two algorithms, i.e. the one that yields the throughput maximizing allocation and the one that yields the completion time minimizing allocation, shall only differ in the termination condition. Throughout the time interval in which a constant power transmission until the end is not feasible, whether the end is defined by a deadline or end of a packet, the power allocation shall be identical. Consequently, the two algorithms work identically until the termination step is reached.

The throughput maximization policy terminates at a certain time in contrast to the completion time minimization policy ending when a certain number of bits have been transmitted. Therefore the definition of the feasible transmission interval $\mathbf{P}[n]$ in (6)-(8) needs to be modified as

$$p_0[n] = \frac{\sum_{k=0}^{n-1} E_k}{s_n}, \quad p_{max}[n] = \frac{\sum_{k=0}^{n} E_k - E_{max}}{s_n}, 0, \qquad p_0[n].s_n < B \qquad (14)$$

$$\mathbf{P}[n] = \{p \,|\, p_{max}[n] \leq p \leq p_0[n]\}, \qquad\qquad p_0[n].s_n < B \qquad (15)$$

$$\mathbf{P}[n_{max}] = p_0[n_{max}], \qquad\qquad p_0[n].s_{n_{max}} = B \qquad (16)$$

where (16) suggests creating a virtual arrival point at $s_{n_{max}}$ that corresponds to the point for which a constant power transmission with the total energy harvested, ignoring energy constraints, transmits all bits. This is the adapted version of the virtual arrival point approach in (8), and can be interpreted as a candidate point for end of transmission that is selected if found feasible. The point $s_{n_{max}}$ lies between the two arrivals for which $p_0[n_i].s_{n_i} < B < p_0[n_{i+1}].s_{n_{i+1}}$ holds, and can be found by solving the equation

$$B = s_{n_{max}}.r\left(\frac{\sum_{k=0}^{n_i} E_k}{s_{n_i}}\right), \qquad s_{n_i} < s_{n_{max}} \qquad (17)$$

With the updated parameters we now provide the modified algorithm that yields the minimum transmission completion time, and prove that the algorithm yields the optimal solution.

### *Transmission Completion Time Minimization Algorithm, A2*

1) Find the upper bound on the length of the first constant transmission using 9. If $n_{ub} = n_{max}$ transmit with constant power $(\sum_{k=0}^{n_{max}} E_k)/s_{n_{max}}$ until end of transmission.

2) Determine whether the next power interval $\mathbf{P}[n_{ub} + 1]$ falls below or above $\bigcap_{k=0}^{n_{ub}} \mathbf{P}[k]$.

3) If $\mathbf{P}[n_{ub} + 1] > \bigcap_{k=0}^{n_{ub}} \mathbf{P}[k]$, transmit with

$$i_1 = s_{n_1}, \quad p_1 = p_0[n_1], \quad \text{where } n_1 = \max\{n \mid p_0[n] \in \bigcap_{k=0}^{n} \mathbf{P}[k]\},$$

If $\mathbf{P}[n_{ub} + 1] < \bigcap_{k=0}^{n_{ub}} \mathbf{P}[k]$, transmit with

$$i_1 = s_{n_1}, \quad p_1 = p_{max}[n_1], \quad \text{where } n_1 = \max\{n \mid p_{max}[n] \in \bigcap_{k=0}^{n} \mathbf{P}[k]\}.$$

4) Repeat algorithm for the shifted problem with modified parameters

$$E'_0 = \sum_{k=0}^{n_1} E_k - i_1.p_1, \quad B' = B - r(p_1).s_{n_1}, \quad n'_{max} = n_{max} - i_1,$$

$$s'_n = s_{n+n_1} - i_1, \qquad E'_n = E_{n+n_1}, \qquad n = 1, ..., n'_{max}. \tag{18}$$

*Theorem 3:* Algorithm A2 gives the optimal power allocation scheme for the transmission completion time minimization problem.

*Proof:* We shall make use of the relation between the two problems to simplify this proof. If the suggested power allocation with completion time $T^*$ is identical to the throughput maximizing allocation with a time constraint $T^*$, we can state that no other allocation with time constraint less than $T^*$ can depart $B$ bits, and therefore the allocation is optimal in time minimization.

This algorithm appears to differ from the throughput maximizing algorithm presented in Section III at only two points: (14)-(16) and (18). However, (18) only affects the termination condition through $n_{max}$ in (16). Consequently the two allocations found using the two algorithms are identical until either one reaches a termination step. The termination step of the time minimization algorithm is reached when there exists a feasible constant power transmission step that departs all the remaining bits by $T^*$. The presence of this step implies that the matching throughput maximization problem with deadline $T^*$ would at the same time have the same feasible constant power transmission opportunity and terminate. Conversely if there does not exist a feasible last step departing all remaining bits, then $T^*$ is unreachable by a constant power transmission and the throughput maximization problem also fails to terminate. Thus we can state that the matching algorithms terminating at the same time $T^*$ would reach the termination step at the same time.

The termination step on the other hand suggests a constant power transmission until the end of transmission at $T^*$. As the previous power policies for the two algorithms are identical, the energies left for the last step are equal. Therefore for any termination step decided by the time minimizing algorithm ending at $T^*$, the throughput maximizing policy with the deadline $T^*$ is forced to take the same step mainly because a constant power transmission is feasible with the available energy. As a result, this algorithm yields a power allocation that is identical to a maximum throughput solution, and is therefore optimal by Theorem 2. ∎

## V. Simulation Results

In this section, we present simulation results to demonstrate the behavior and the performance of the algorithms. First, we have a sample simulation run of the algorithm presented in Section III-B. The simulation is performed for a node with $E_{max} = 10$ units with energy arrivals of $E_n = \{2, 1, 6, 4, 8, 1\}$ units at times $s_n = \{0, 2, 4, 5, 7, 11\}$ respectively. The energy arrivals are shown in Figure 5(a). The energy tunnel of this system is shown in Figure 5(b) along with the result of the throughput maximization algorithm for a deadline constraint $T = 12$.

At the first step of the algorithm, the sets $p_0[n]$ and $r_{max}[n]$ are calculated and the corresponding intervals $\mathbf{P}[n]$ are determined. These intervals are displayed in Figure 5(c) with vertical lines. The upper bound on $n_1$ is determined to be $n_{ub} = 3$ as the third interval $\mathbf{P}[4]$ falls outside $\mathbf{P}[1] \cap \mathbf{P}[2] \cap \mathbf{P}[3]$. Based on the position of $\mathbf{P}[4]$, the longest feasible transmission with power in $p_0[n]$ is picked, and the first step of the algorithm is decided as $p_1$ with length $s_{n_1}$. When new intervals for the shifted problem are calculated beyond $s_{n_1}$, an opposite position for $\mathbf{P}[n_{ub} + 1]$ is observed and a transmission within $p_{max}[n]$ is picked instead. Finally, for the next shifted problem, the constant power transmission is found to be feasible and the algorithm is terminated.

Observe that no other power allocation policy within the energy feasible tunnel can perform better than this. This can be proved by stating that any different power curve reaching from the origin to the end point at $T = 12$ must cross this line at least once, at which point it has spent the same amount of energy while departing strictly less bits.

Assuming the same energy arrival scenario, we next observe the performance of the algorithm presented in Section III-B and the algorithm presented in Section IV for a range of deadline and packet size parameters in Figure 6. As predicted by Theorem 2, the curves match up perfectly, verifying the relationship between the two problems. A point $(B_1, T_1)$ on this curve corresponds to a power allocation that solves both of the problems in consideration simultaneously: The throughput maximization problem for a given deadline $T_1$ is solved by departing a maximum of $B_1$ bits, while the transmission completion time minimization problem for a given packet size of $B_1$ bits is solved by completing the transmission in $T_1$ with an identical power allocation policy. Another noteworthy observation from Figure 6 is the strictly increasing nature of both of the problems with respect to their parameters. As a consequence, more time is required to depart a longer packet, or similarly more bits can be transmitted when given a more lenient deadline.

After observing the behavior of the algorithms on a smaller time scale, we simulate for longer realizations in order to evaluate the average long term performance of the algorithms. Assuming a battery capacity $E_{max} = 100$ we generate energy arrivals randomly with energy packet size distributed uniformly in $[0, E_{max}]$ and inter-arrival times distributed exponentially with an average of 5 seconds. In this setting, Figure 7 compares our optimal power allocation algorithm for the throughput maximization problem with two alternative power allocations for a deadline of $T = 10000$ sec. First corresponds to transmission without regard to any battery or arrival constraints. This is the performance of a "traditional" transmitter with no energy harvesting, and is presented as an upper bound for our model where we are bound to conform to energy feasibility and battery constraints. Second is an alternative algorithmic approach named the on-off algorithm. It is based on the fact that for a strictly concave power-rate relationship, constant power transmission is the most efficient. The policy is such that, the transmitter operates with constant power when energy is available, and shuts off when energy is depleted. The constant power level is determined from the average energy arrival rate. As seen in Figure 7, the energy constraints of the problem result in a performance loss with respect to having perpetual energy. However, a major portion of this loss can be recovered with the incorporation of arrival information and use of variable power transmission. The optimal offline algorithm we present provides this remedy, and significantly outperforms the greedy on-off algorithm.

Throughout further simulation runs involving various battery capacity and arrival statistics within a larger number of randomly generated arrivals, we observed that the optimal offline algorithms provided improvements over simpler policies such as the on-off transmission algorithm suggested above. The improvements were especially notable for the cases of relatively small battery capacity $E_{max}$, large energy arrival amounts $E_n$ and large standard deviation on the arrival process. Therefore, it can be stated that employing optimal power allocation is more beneficial for systems with limited battery capacity or energy arrivals with large variations.

## VI. Conclusion

In this paper, we have solved the short-term throughput maximization problem for a link with an energy harvesting transmitter, and limited energy storage capacity. Furthermore, we have shown that obtaining the maximum amount of data transferred by a given deadline, is also equivalent to solving for the minimum completion time (the deadline) given this amount of data. We have proposed the algorithms that yield the optimal solution of both problems and proved their optimality.

The findings of this paper provide insight to developing optimal transmission policies for nodes that have some notion of when and how much energy they can harvest. Future directions include developing online power allocation algorithms for causal systems or systems with stochastic future energy harvests. There are also plenty of possible extensions for multiterminal system models from the single link model considered in this paper.

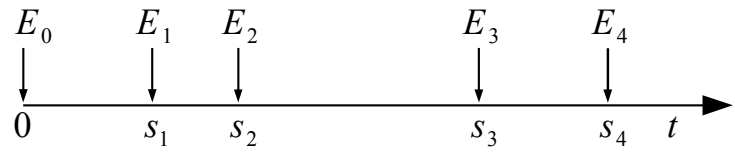

Fig. 1: Energy harvesting model.

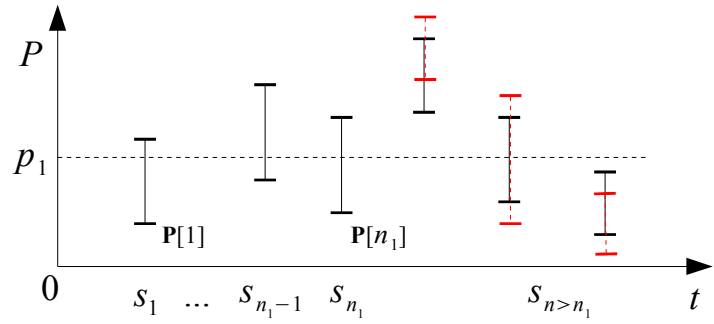

Fig. 2: Pictorial depiction of the behavior of feasible power ranges for future arrivals.

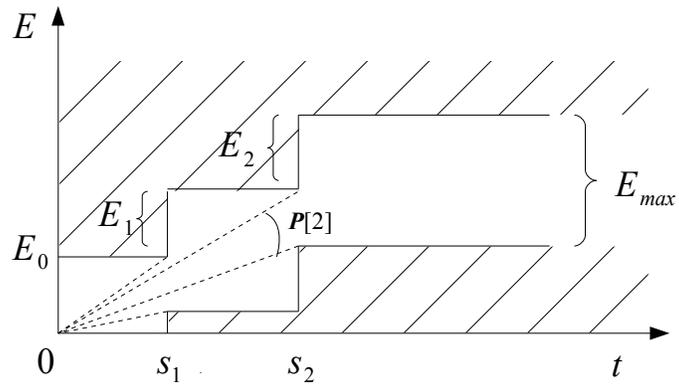

Fig. 3: The feasible energy tunnel.

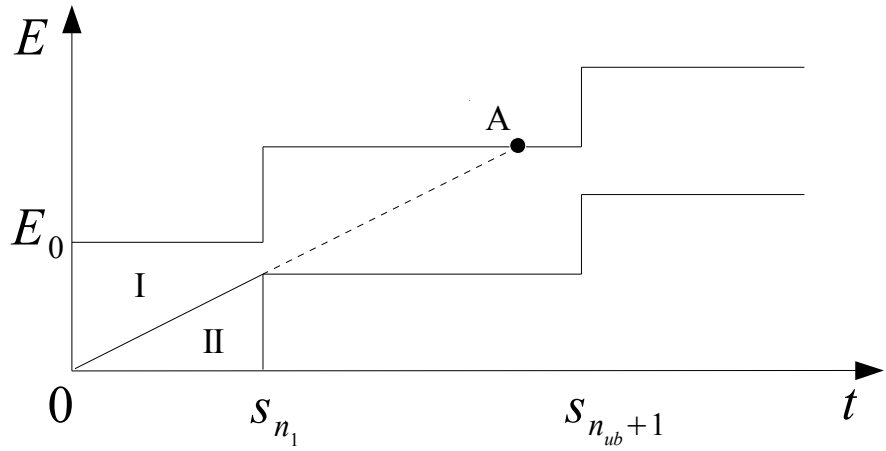

(a) $E_{max}$ case

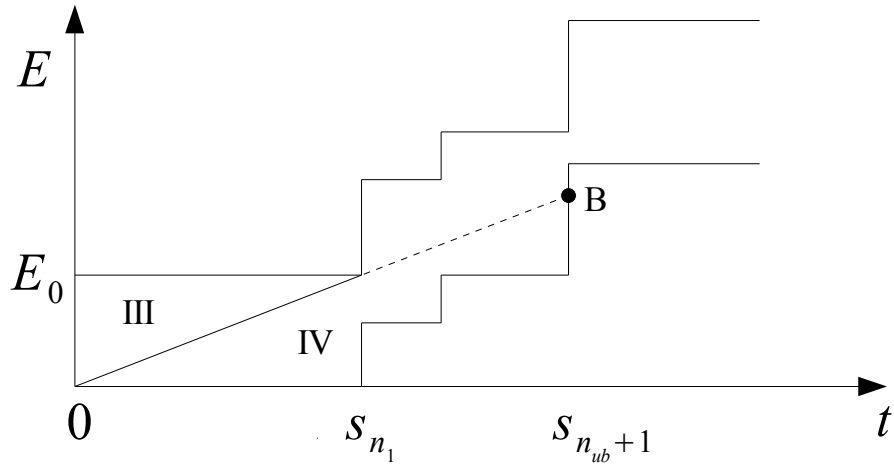

(b) $E_0$ case

Fig. 4: The optimal first step and suboptimal regions.

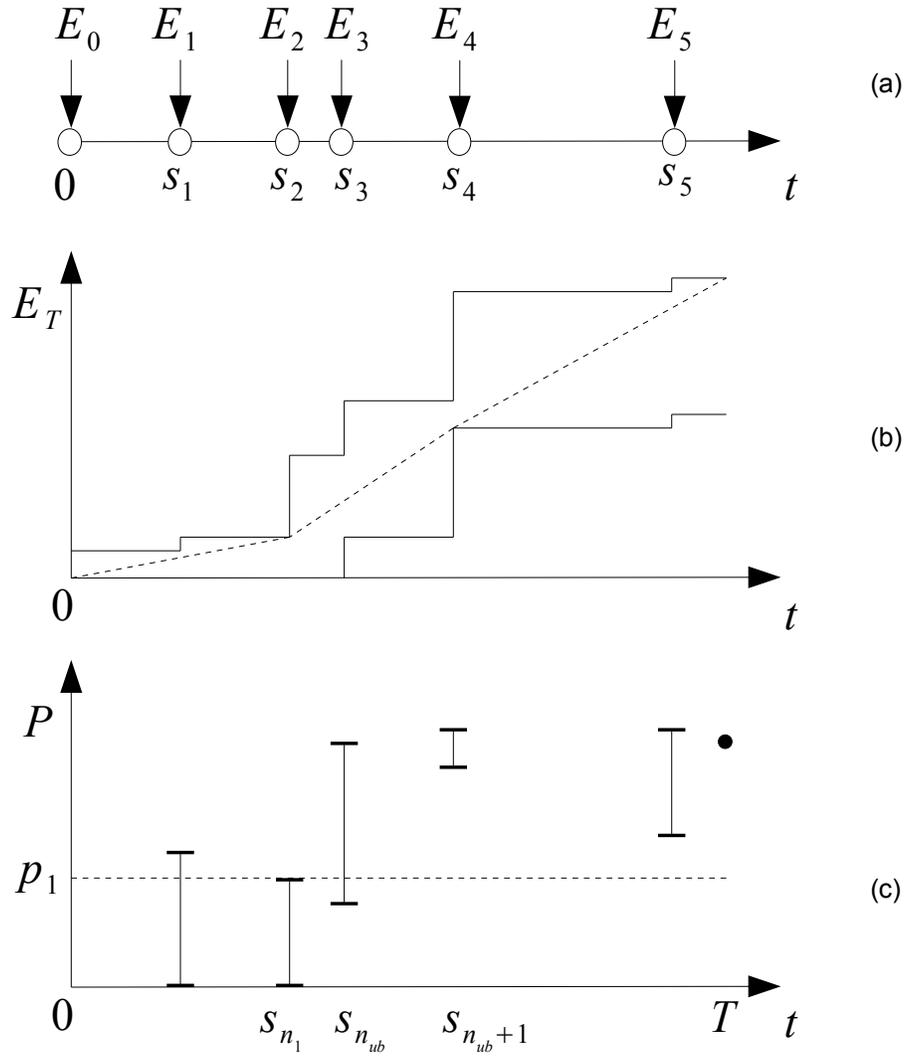

Fig. 5: Arrival scenario and simulation results for the sample run.

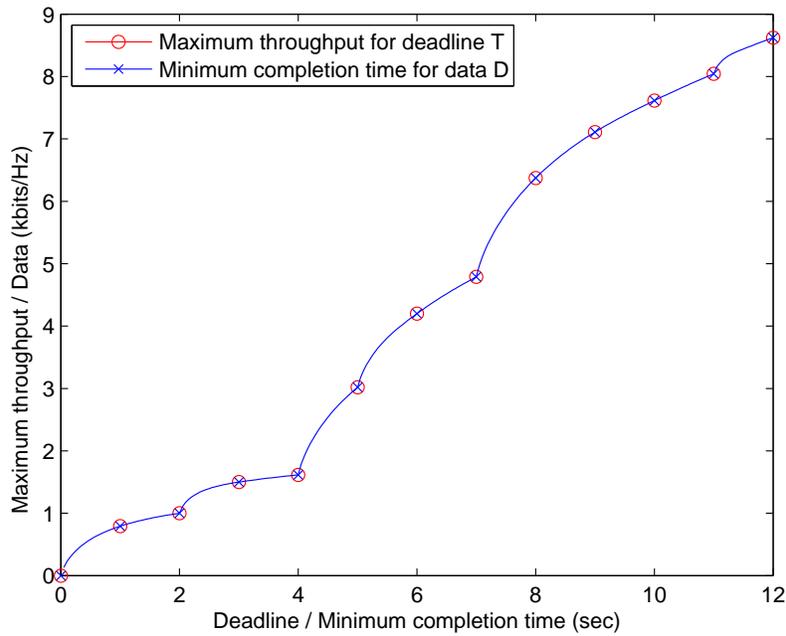

Fig. 6: Overlayed plot of throughput by deadline and completion time by packet size for the energy harvest scenario in Figure 5.

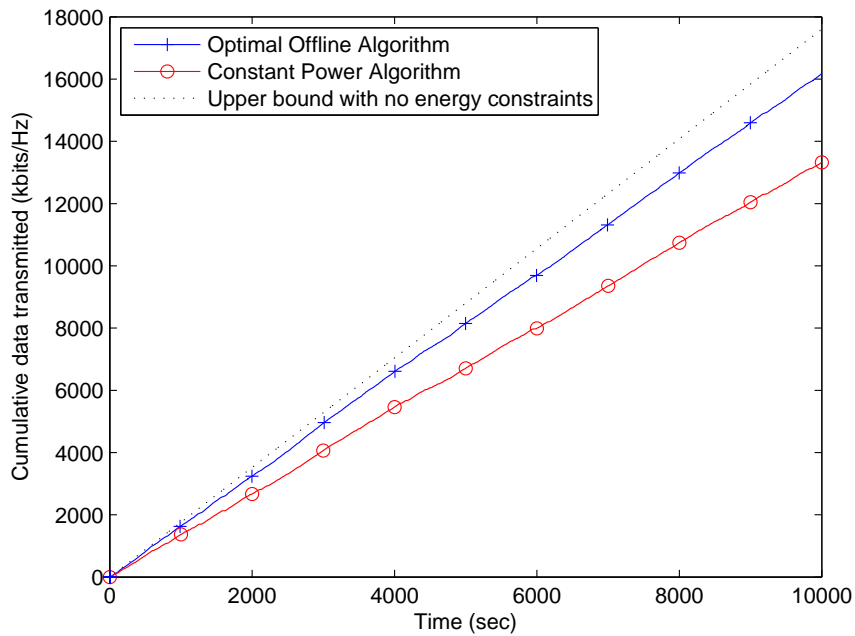

Fig. 7: Performance comparison of transmission policies for a transmission of $T = 10000s$ with $E_{max} = 100$, $E_n$ distributed uniformly in $[0, E_{max}]$, and interarrival times $s_n - s_{n-1}$ distributed exponentially with a mean of 5 seconds.